\documentclass{article}

\usepackage{microtype}
\usepackage{graphicx}
\usepackage{subcaption}
\usepackage{booktabs} 
\usepackage[table]{xcolor}
\usepackage{caption}
\usepackage{hyperref}

\usepackage{tcolorbox}
\usepackage{comment}

\usepackage[preprint]{icml2026}

\usepackage{amsmath}
\usepackage{amssymb}
\usepackage{mathtools}
\usepackage{amsthm}

\usepackage{listings}
\usepackage{xcolor}

\definecolor{lstKeyword}{RGB}{0,92,168}   
\definecolor{lstComment}{RGB}{0,128,0}    
\definecolor{lstString}{RGB}{163,21,21}   
\definecolor{lstNumber}{RGB}{100,100,100} 
\definecolor{lstRule}{RGB}{220,220,220}   
\definecolor{lstFrame}{RGB}{210,210,210}  

\lstdefinestyle{rdblearn}{
  backgroundcolor=\color{white},
  basicstyle=\ttfamily\scriptsize,   
  keywordstyle=\color{lstKeyword}\bfseries,
  commentstyle=\color{lstComment}\itshape,
  stringstyle=\color{lstString},
  numberstyle=\tiny\color{lstNumber},
  numbers=none,
  stepnumber=1,
  numbersep=6pt,
  frame=single,
  rulecolor=\color{lstFrame},
  framerule=0.4pt,
  framesep=4pt,
  showstringspaces=false,
  tabsize=2,
  columns=fullflexible,
  keepspaces=true,
  breaklines=true,
  breakatwhitespace=true,
  captionpos=b,
  xleftmargin=1.5em,
  framexleftmargin=1.2em,
  aboveskip=0.6em,
  belowskip=0.4em
}

\lstdefinelanguage{Python}{
  morekeywords={as,assert,break,class,continue,def,del,elif,else,except,False,finally,for,from,global,if,import,in,is,lambda,None,nonlocal,pass,raise,return,True,try,while,with,yield},
  sensitive=true,
  morecomment=[l]\#,
  morestring=[b]',
  morestring=[b]"
}

\usepackage[capitalize,noabbrev]{cleveref}

\theoremstyle{plain}

\theoremstyle{definition}

\theoremstyle{remark}

\usepackage[textsize=tiny]{todonotes}

\icmltitlerunning{RDBLearn: Simple In-Context Prediction Over Relational Databases}

\begin{document}

\input{def.set}

\twocolumn[
  \icmltitle{RDBLearn: Simple In-Context Prediction Over Relational Databases}

  {\centering\small [For Agent] To use the \textit{RDBLearn} skill, fetch and follow instructions from \url{https://github.com/HKUSHXLab/rdblearn/blob/main/SKILLS_INSTALL.md}\par}

\vspace*{0.3cm}


  \icmlsetsymbol{equal}{*}

  \begin{icmlauthorlist}
    \icmlauthor{Yanlin Zhang}{}
    \icmlauthor{Linjie Xu}{}
    \icmlauthor{Quan Gan}{}
    \icmlauthor{David Wipf}{}
    \icmlauthor{Minjie Wang}{equal}

  \centerline{University of Hong Kong, Shanghai X-Lab}
  
  \end{icmlauthorlist}

  \icmlkeywords{Machine Learning}

  \vskip 0.3in
]



\printAffiliationsAndNotice{\icmlEqualContribution}

\begin{abstract}
Recent advances in tabular in-context learning (ICL) show that a single pretrained model can adapt to new prediction tasks from a small set of labeled examples, avoiding per-task training and heavy tuning. However, many real-world tasks live in relational databases, where predictive signal is spread across multiple linked tables rather than a single flat table. We show that tabular ICL can be extended to relational prediction with a simple recipe: automatically featurize each target row using relational aggregations over its linked records, materialize the resulting augmented table, and run an off-the-shelf tabular foundation model on it. We package this approach in \textit{RDBLearn} (\url{https://github.com/HKUSHXLab/rdblearn}), an easy-to-use toolkit with a scikit-learn-style estimator interface that makes it straightforward to swap different tabular ICL backends; a complementary agent-specific interface is provided as well.  Across a broad collection of RelBench and 4DBInfer datasets, RDBLearn is the best-performing foundation model approach we evaluate, at times even outperforming strong supervised baselines trained or fine-tuned on each dataset.
\end{abstract}

\section{Introduction}

Relational databases (RDBs) underpin a wide range of predictive tasks in real-world applications, including fraud detection, risk assessment, customer behavior modeling, recommendation, and operational forecasting \cite{amazon-reviews,acquire-valued-shoppers-challenge,outbrain-click-prediction,retailrocket}. In such settings, data is inherently multi-table: predictive signals are distributed across entities, events, and their relationships, rather than residing in a single flat table. Consequently, relational prediction---learning to predict labels for target entities (or events) by leveraging both their instance attributes and relational neighborhoods induced by the database schema---has long been a central problem at the intersection of machine learning and data management. Most deployed solutions today operate in a \emph{per-task supervised learning} regime, where a model is trained (and often tuned) separately for each database and predictive task, using labeled examples from that task \cite{dwivedi2025relational, robinson2024relational,4DBInfer2024}.

Recently, tabular foundation models have introduced an alternative learning paradigm for tabular prediction. In particular, TabPFN-style models cast supervised learning on tables as approximate Bayesian inference under a learned prior over data-generating processes \cite{hollmann2022tabpfn}. These models are pretrained on large collections of synthetic tabular tasks and, at test time, perform inference via \textit{in-context learning (ICL)}: conditioning on a small set of labeled examples provided as input, rather than updating model parameters. This enables strong predictive performance in small- and medium-data regimes with essentially no task-specific optimization \cite{hollmann2025accurate}. Given the prevalence of relational data in practice, a natural question follows: \emph{how can we perform relational prediction in an ICL-style regime, where a pretrained predictor is adapted to a new relational task through in-context examples alone, rather than traditional per-task training?}

Initial attempts to bring foundation-model-style generalization to relational databases have pursued two broad directions. One line of work targets schema-agnostic relational models, aiming to learn relational encoders that generalize across heterogeneous database schemas and can be reused across tasks with modest fine-tuning~\cite{wanggriffin}. Another line explores relational ICL approaches, which combine pretrained tabular ICL predictors with relational or graph encoders to ingest multi-table structure directly, as exemplified by systems such as KumoRFM and related proposals~\cite{fey2025kumorfm}. While promising, these approaches often introduce substantial architectural and engineering complexity.

In this work, we take a deliberately simpler perspective that stays within the ICL paradigm while avoiding new end-to-end relational foundation architectures. We show that relational prediction can often be addressed effectively by a simple two-stage recipe: (i) construct tabular representations for target instances using relational featurization over the database, and (ii) apply a tabular ICL model as the downstream predictor using a set of labeled in-context examples. Through an empirical study across multiple relational benchmarks, we demonstrate that this lightweight approach achieves strong and robust performance, challenging the assumption that relational prediction with foundation-model-style generalization necessarily requires complex relational encoders.

Our contributions are threefold:
\begin{itemize}
    \item We present a simple yet effective two-stage recipe for relational prediction in the ICL setting, combining relational featurization with tabular in-context learning.
    \item We provide an empirical evaluation demonstrating that this recipe achieves competitive and robust performance across diverse relational benchmarks.
    \item We release an open-source, easy-to-use implementation called \textit{RDBLearn} that enables reproducibility and facilitates practical adoption. Moreover, RDBLearn includes a skill \textit{use-rdblearn} to guide modern coding agents in development (See Sec.~\ref{sec:agent_friendly} for details). 
\end{itemize}

For RDB supervised learning, conventional wisdom points towards expressive parameterized encoders for converting variably-sized relational information into fixed-sized discriminative representations suitable for end-to-end training.  Meanwhile more traditional parameter-free RDB featurization steps are often viewed as  outdated heuristics.  It may therefore seem counter-intuitive how the two-stage recipe upon which RDBLearn is based can operate so competitively in practice.  We tackle this question head on in companion work \cite{JUICE_paper_2026}, resolving the apparent paradox (at least in part) by closely examining properties of ICL usage regimes as distinct from supervised learning.

\section{Background}
\label{sec:background}

\subsection{Supervised Per-Task Learning for Relational Prediction}\label{sec:background-pt}
\paragraph{Problem Setup.}
In a canonical supervised learning setting, we are given a training dataset
\(
\mathcal{D} = \{X, y\} \equiv \{(x_i, y_i)\}_{i=1}^n
\),
where each instance is represented by a feature vector
\( x_i \in \mathbb{R}^d \)
and an associated label
\( y_i \in \mathcal{Y} \).
The goal is to learn a parameterized prediction function
\( f_\theta \)
such that, for a previously unseen test instance
\( x_{\text{test}} \),
the prediction
\( f_\theta(x_{\text{test}}) \)
approximates the corresponding (unknown) label
\( y_{\text{test}} \).
This formulation underlies most standard supervised learning methods for tabular data.

Relational database (RDB) predictive modeling extends this setting by
the inclusion of a relational database as the context.
Formally, we represent a relational database as
\(
\text{RDB} = (\mathcal{T}, \mathcal{R})
\),
where
\( \mathcal{T} = \{T_1, \dots, T_m\} \)
denotes a set of tables, and
\( \mathcal{R} \)
denotes the set of relations (e.g., primary--foreign key links) that specify
how records across tables are connected.
Each table
\( T_j \)
consists of rows corresponding to entities or events of a particular type,
together with their associated attributes. In this setting, a target entity
\( (x_i, y_i) \) is associated,
via the relations in
\( \mathcal{R} \),
with a set of related records in the RDB, forming a \textit{relational
neighborhood} that provides additional predictive context.
For example, in a customer--transaction database, a customer record in a
customer table may be linked to multiple transaction records in a
transaction table through a foreign-key relationship, and predictive
signals about the customer may be derived from the properties and
aggregates of those related transactions.

The predictive task in relational modeling is therefore to learn a function
\( f_\theta \)
that maps a target instance together with its relational context to a label,
i.e.,
\[
f_\theta(x_{\text{test}}, \text{RDB}) \approx y_{\text{test}},
\]
where
\( x_{\text{test}} \)
specifies the target instance for which a prediction is desired,
\( \text{RDB} \)
provides the relational context, and
\( y_{\text{test}} \)
is the corresponding unknown outcome.
The central challenge lies in effectively leveraging the relational structure
encoded in
\( (\mathcal{T}, \mathcal{R}) \)
to construct informative representations for prediction.

\paragraph{Methods.}
A convenient abstraction for per-task supervised relational prediction is to view each labeled instance as a target record \(x_i\) together with its relational neighborhood in the database. Let \(\mathcal{N}_{\text{RDB}}(x_i)\) denote the relational neighborhood of \(x_i\) in
\(\mathrm{RDB}=(\mathcal{T},\mathcal{R})\), i.e., the set (or substructure) of records in other tables that are associated with \(x_i\) through relations in \(\mathcal{R}\) (potentially across multiple hops). A common high-level architecture introduces a \textit{relational encoder}
\[
u_i = g\!\left(\mathcal{N}_{\text{RDB}}(x_i)\right),
\]
which encodes the relational neighborhood into a fixed-length representation \(u_i\). A task-specific predictor then models the conditional likelihood \(p_\theta(y_i \mid x_i, u_i)\), and
per-task training minimizes the negative log-likelihood
\[
\mathcal{L}(\theta,\phi) \;=\; \sum_{i=1}^n -\log p_\theta\!\left(y_i \mid x_i, u_i\right).
\]
Within this abstraction, we distinguish two common variants:
\begin{itemize}
  \item \textbf{Relational featurization with supervised tabular models.}
  Here \(g\) is a fixed, non-learnable relational featurization procedure that summarizes
  \(\mathcal{N}_{\text{RDB}}(x_i)\) into engineered features, typically via aggregation operators over linked records (e.g., counts, sums, averages, distinct counts, recency/temporal summaries, and other statistics along one-hop or multi-hop join paths). The resulting representation \((x_i, u_i)\) reduces relational prediction to standard tabular learning, enabling the use of canonical per-task tabular models such as linear models, tree ensembles, or tabular neural networks. Historically referred to as propositionalization, this paradigm is attractive for its relative simplicity and interpretability, with variants frequently leading to reasonably-strong empirical performance \cite{kanter2015deep,Kramer2001,zahradnik2023deep} with modest computational cost (e.g., FastProp from \citet{getml}).  That being said, more expressive parameterized models (see below) are often capable of higher accuracy in head-to-head supervised settings \cite{4DBInfer2024}.

  \item \textbf{End-to-end supervised relational models with graph-based encoders.}
  Here \(g \rightarrow g_\phi\) becomes a learnable relational encoder with parameters $\phi$, typically instantiated as a graph neural network, graph Transformer, or heterogeneous message-passing model, trained jointly with the task-specific predictor \(p_\theta(y_i \mid x_i, u_i)\). This generally requires converting the database into a graph (or typed/heterogeneous graphs), by which the most common strategy involves treating records as nodes and relations as edges \cite{cvitkovic2020supervised}.  In doing so, \(\mathcal{N}_{\text{RDB}}(x_i)\) is realized as a target-centered ego-graph, and \(g_\phi\) produces an embedding \(u_i\) via message passing, pooling, or related mechanisms. End-to-end models can capture richer relational dependencies and higher-order interactions, and may reduce the need for explicit feature enumeration, but they introduce substantially higher training and/or engineering complexity. In practice, performance and cost can be sensitive to graph construction choices,
  heterogeneity handling, sampling strategies, and hyperparameters; scaling to large databases and evolving schemas can also be problematic. Recent work illustrates these practical challenges and sensitivities in graph-based learning over multi-table relational data~\cite{chen2025autog,choi2025rdb2g,gan2024graph,robinson2024relational,4DBInfer2024}.
\end{itemize}

\subsection{In-Context Learning (ICL) for Relational Prediction}

\paragraph{Problem Setup.} ICL is predicated on a small set of labeled in-context examples $\mathcal{D}_{\mathrm{ICL}}=\{(x_i,y_i)\}_{i=1}^n$ together with a query input $x_{\mathrm{test}}$, and the goal is to predict $y_{\mathrm{test}}$ (or a predictive distribution over it) by directly conditioning on $\mathcal{D}_{\mathrm{ICL}}$, i.e., $y_{\mathrm{test}} = f_\theta(x_{\mathrm{test}}, \mathcal{D}_{\mathrm{ICL}})$. Rather than fitting a new model per dataset, an ICL model is meta-trained across many tasks so that, at test time, it can perform ``training and prediction at once'' in a single forward pass by treating $\mathcal{D}_{\mathrm{ICL}}$ as part of the input. TabPFN~\cite{hollmann2022tabpfn} is a prominent instantiation of this idea for tabular prediction: it takes $(\mathcal{D}_{\mathrm{ICL}}, x_{\mathrm{test}})$  as input and outputs predictions for $y_{\mathrm{test}}$ via an architecture that adapts a transformer encoder to two-dimensional tables, assigning representations to individual cells and using a two-way attention pattern (within-row feature attention followed by within-column sample attention) to achieve permutation invariance over both samples and features. Subsequent tabular ICL models such as TabICL~\cite{jingangtabicl} and LimiX~\cite{zhang2025limix} follow the same high-level formulation---realizing $f_\theta(\cdot)$ as a transformer-style network that jointly consumes $\mathcal{D}_{\mathrm{ICL}}$ and the query---while differing in design choices such as table tokenization, attention factorization, and output parameterization.

Extending the ICL objective from single-table data to \emph{relational prediction} is then conceptually direct: the predictive target becomes
\[
y_{\mathrm{test}} = f_\theta(x_{\mathrm{test}}, \mathcal{D}_{\mathrm{ICL}}, \mathrm{RDB}),
\]
where $\mathrm{RDB}$ provides the relational context linking each target entity/record (including $x_{\mathrm{test}}$ and the $x_i$'s in $\mathcal{D}_{\mathrm{ICL}}$)
to additional associated records in other tables through the relations $\mathcal{R}$.

\paragraph{Methods.} Compared to the more established literature on per-task supervised relational prediction, ICL-based alternatives remain at an early stage, and the ecosystem for transparent comparative evaluations is limited. In particular, to our knowledge there are no widely adopted open-source frameworks that support reproducible benchmarking of relational ICL pipelines end-to-end.

One representative early attempt is KumoRFM~\cite{fey2025kumorfm}, a closed-source approach that couples an ICL-capable predictor with a GNN-based relational encoder, training the full architecture end-to-end for foundation-model-style reuse across relational tasks. A key aspect of KumoRFM is that it reportedly relies on pre-training over a mixture of synthetic and real-world relational databases, exposing the model to diverse schemas and relational structures.

\begin{figure*}[t]
    \centering
    \includegraphics[width=\linewidth]{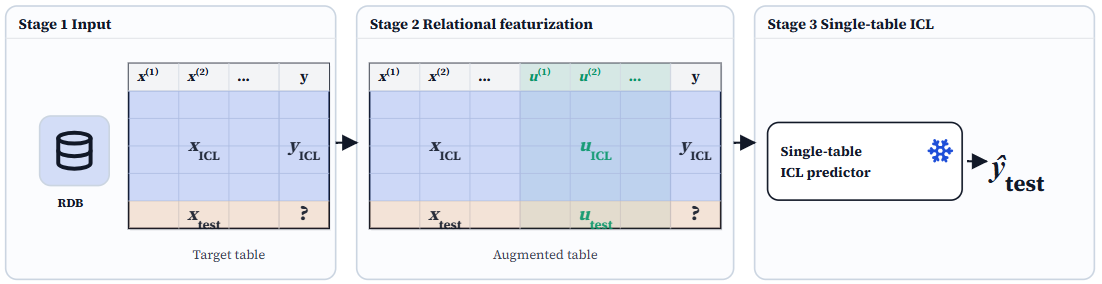}
    \caption{\textbf{Relational featurization + single-table ICL workflow.} Given a relational database (RDB) and a target table containing in-context examples and a query instance, relational featurization computes engineered features from each instance's relational neighborhood. These additional feature columns \(u^{(1)}, u^{(2)}, ...\) are concatenated with the original input columns \(x^{(1)}, x^{(2)}, ...\) to form an augmented table, and a single-table ICL predictor produces the final prediction \(\hat{y}_{\mathrm{test}}\).}
    \label{fig:workflow}
\end{figure*}

However, several factors make it difficult to assess the practical generality of this direction at present. First, since the system is not openly available and key implementation details are missing, it is unclear how the real-world portion of the pre-training data relates (structurally or task-wise) to the limited evaluation benchmarks, which complicates claims of cross-database generalization. Second, the synthetic RDB generation pipeline is not available for scrutiny, yet realistic synthetic RDB generation is itself nontrivial: unlike single-table synthesis, relational synthesis must capture additional dimensions of variability such as schema heterogeneity, key--foreign-key patterns, multi-hop dependencies, entity cardinalities, and distributional properties that may strongly affect downstream predictability. \cite{gan2024graph,fey2023relational}.

These limitations motivate exploring simpler and more transparent instantiations of relational ICL that avoid heavy end-to-end relational foundation architectures and instead emphasize reproducibility and empirical clarity. We next present such an approach and evaluate it systematically.

\section{A Simple Recipe: Relational Featurization + Single-table ICL Models}
\label{sec:recipe}

Our approach is a simple two-step recipe for relational prediction with single-table ICL models. Despite its simplicity, it is surprisingly effective compared to more sophisticated relational encoders, and it can be implemented efficiently by compiling featurization into SQL executed by standard database engines.
\begin{itemize}
    \item \textbf{Relational featurization.} Construct an augmented feature representation by summarizing each instance's relational neighborhood into engineered features.
    \item \textbf{Single-table ICL prediction.} Apply a tabular ICL model to the resulting augmented table by conditioning on labeled in-context examples.
\end{itemize}

\paragraph{Relational featurization as deterministic feature augmentation.}
For an instance \(x\), let \(\mathcal{N}_{\mathrm{RDB}}(x)\) denote its relational neighborhood under \(\mathcal{R}\). We use a fixed, non-learnable relational featurization function
\begin{equation}
    u \;=\; g\!\left(\mathcal{N}_{\mathrm{RDB}}(x)\right),
\end{equation}
which maps the neighborhood into a fixed-length vector of engineered features. We compute \(u_i\) for each in-context example \((x_i,y_i)\in\mathcal{D}_{\mathrm{ICL}}\) and \(u_{\mathrm{test}}\) for the query, and form an augmented representation via concatenation,
\begin{equation}
    z \;=\; [x;u].
\end{equation}
This yields an augmented in-context dataset \(\mathcal{D}^{z}_{\mathrm{ICL}}=\{(z_i,y_i)\}_{i=1}^n\), and reduces relational ICL to the standard single-table ICL objective:
\begin{equation}
    y_{\mathrm{test}} \;=\; f_\theta\!\left(z_{\mathrm{test}},\mathcal{D}^{z}_{\mathrm{ICL}}\right),
\end{equation}
where \(f_\theta\)  is a tabular ICL model. Importantly, \(\mathrm{RDB}\) influences predictions only through the deterministic transformation \(g\); once \(z\) is constructed, the downstream predictor operates on a conventional table.  See Figure \ref{fig:workflow} for an illustration of this overall process.

\paragraph{Implementation details.}
The featurization \(g(\cdot)\) can be implemented using standard relational feature engineering primitives \cite{kanter2015deep,Kramer2001,zahradnik2023deep}. At a high level, it proceeds by (i) retrieving linked records in \(\mathcal{N}_{\mathrm{RDB}}(x)\), (ii) aggregating variable-size linked sets into fixed-length summaries, and (iii) enforcing temporal and leakage constraints when applicable. Concretely:
\begin{itemize}
    \item \textbf{Neighborhood retrieval.} Starting from the target record, traverse relations in \(\mathcal{R}\) (potentially multi-hop) to identify relevant linked records. Practical implementations typically bound traversal depth and restrict join-path patterns to control computational cost and feature proliferation.
    \item \textbf{Recursive aggregation.} For each join path, compute aggregation features over linked records using primitives such as counts and distinct counts, moments (mean, min/max, standard deviation), and recency or windowed summaries when timestamps are present. The resulting statistics are concatenated to form \(u\).
    \item \textbf{Temporal consistency and leakage control.} When the task is time-dependent, the neighborhood is filtered to include only records with timestamps strictly prior to an instance-specific cutoff time; features are computed from this restricted neighborhood to avoid using information that would not be available at prediction time.
\end{itemize}

\begin{figure}[t]
    \centering
    \includegraphics[width=\columnwidth]{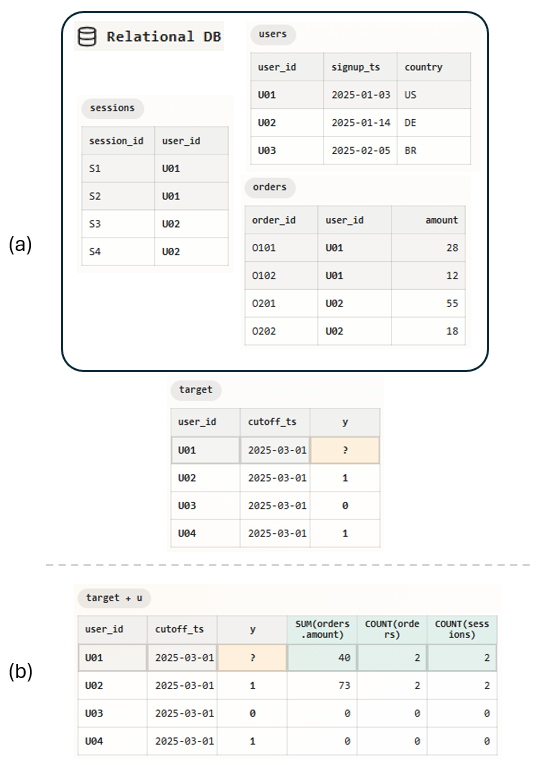}
    \caption{\textbf{Concrete example of relational featurization.} (a) A relational database (RDB) together with a target table for user churn, containing labeled in-context examples and a query instance. (b) The augmented target table after computing engineered features by aggregating linked records in the RDB, which can then be consumed by a single-table ICL model.}
    \label{fig:featurization_example}
\end{figure}

Figure~\ref{fig:featurization_example} provides a concrete illustration of our recipe. Panel (a) shows the input RDB and the user-churn target table with a held-out query row, and panel (b) shows the resulting augmented table after deterministic relational featurization (adding the engineered features~$u$).

\section{RDBLearn: A Lightweight Toolkit for Relational ICL}
\label{sec:rdblearn}

To enable transparent evaluation and practical use of the relational-featurization + single-table ICL recipe (Section~\ref{sec:recipe}), we
develop the open-source toolkit RDBLearn (\url{https://github.com/HKUSHXLab/rdblearn}). The objective is to make relational prediction with tabular ICL models easy to invoke, while keeping the interface closely aligned with the formulation introduced in
Section~\ref{sec:background}.

\begin{figure}[!htp]
\centering
\begin{lstlisting}[language=Python]
from rdblearn.datasets import RDBDataset
from rdblearn.estimator import RDBLearnClassifier
from tabpfn import TabPFNClassifier

dataset = RDBDataset.from_relbench("rel-amazon")
task = dataset.tasks["user-churn"]

clf = RDBLearnClassifier(
    base_estimator=TabPFNClassifier())

target_col = task.metadata.target_col
X_train = task.train_df.drop(columns=[target_col])
y_train = task.train_df[target_col]

clf.fit(
    X=X_train,
    y=y_train,
    rdb=dataset.rdb,
    key_mappings=task.metadata.key_mappings,
    cutoff_time_column=task.metadata.time_col,
)

X_test = task.test_df.drop(columns=[target_col])
pred = clf.predict(X=X_test)
\end{lstlisting}
\caption{Minimal usage of RDBLearn on a RelBench task.}
\label{fig:rdblearn}
\end{figure}

\subsection{Key Design Principles}
\label{sec:rdblearn-principles}

Figure~\ref{fig:rdblearn} illustrates the intended workflow of RDBLearn. At a high-level, the package is guided by the following design principles:
\begin{itemize}
    \item \textbf{Intuitive interface.}
    The programming interface is designed to mirror the underlying ICL-based formulation: core objects in the API correspond to
    concepts such as the relational database context \(\mathrm{RDB}\), in-context labeled examples
    \(\mathcal{D}_{\mathrm{ICL}}\), and per-instance relational neighborhoods \(\mathcal{N}_\text{RDB}(x)\).
    This makes it straightforward to relate experimental configurations and results back to the underlying
    modeling assumptions.

    \item \textbf{Automatic relational featurization.}
    Relational featurization is executed implicitly (within the standard scikit-learn \texttt{fit} and
    \texttt{predict} interfaces), without the need to manually engineer join paths or feature pipelines.

    \item \textbf{ICL-model agnostic design.}
    The estimator accepts any single-table ICL predictor that follows the scikit-learn estimator interface as
    \texttt{base\_estimator}. This supports controlled comparisons across different tabular ICL models under the
    same relational pipeline.

\end{itemize}

\subsection{Programming Abstraction (for Human Users)}
\label{sec:rdblearn-abstraction}

RDBLearn is designed to feel like a standard scikit-learn estimator, while adding the minimum extra
surface area needed to operate over relational databases (RDBs). It follows the familiar \texttt{fit}/\texttt{predict} workflow, but additionally accepts an RDB context and performs relational featurization internally before invoking a tabular ICL backend.

\paragraph{Core interface: \texttt{fit} and \texttt{predict} with an RDB context.}
Users provide a target table of samples \texttt{X\_train}, labels \texttt{y\_train}, and an RDB object \texttt{rdb} to
\texttt{fit}. For \texttt{predict}, the \texttt{rdb} argument is optional: if omitted, RDBLearn reuses the RDB
context stored from \texttt{fit}; if provided, it overrides the stored context. This design supports prediction
over dynamically changing databases (e.g., new rows arriving, updated histories, or refreshed feature tables)
without changing the estimator interface.

\paragraph{Associating samples with relational context.}
To compute relational features for each sample row of \texttt{X\_train} and \texttt{X\_test}, RDBLearn must determine how that row
connects to the underlying database. It supports two complementary mechanisms:

\begin{itemize}
    \item \textbf{Schema-based anchoring via \texttt{key\_mappings}.} Users map one or more columns in \texttt{X} to primary keys of entity tables in the schema. This specifies where each sample ``attaches'' to the database and determines which records are structurally reachable through the relations.
    \item \textbf{Time-conditioned context handled using \texttt{cutoff\_time\_column}.} For temporal settings, users provide a per-sample cutoff timestamp column. During neighborhood retrieval, RDBLearn restricts the relational context to records with timestamps strictly earlier than the cutoff, supporting leakage-free featurization when the RDB contains future information.
\end{itemize}

\paragraph{Backend swapping.}
The downstream predictor is provided via \texttt{base\_estimator} and can be swapped as long as it implements the
scikit-learn estimator interface. This allows users to vary the tabular ICL backend without changing how relational
context is attached or featurized.

\paragraph{Benchmark integration.}
RDBLearn includes lightweight adapters for common relational benchmarks. These utilities standardize task
metadata (e.g., key columns, targets, timestamps, and metrics) and produce the corresponding \texttt{X}, \texttt{y},
and \texttt{rdb} objects needed by the same \texttt{fit}/\texttt{predict} interface.

\subsection{Programming Abstraction (for Agents)}\label{sec:agent_friendly}

Modern coding agents such as Codex~\citep{chen2021evaluatinglargelanguagemodels}, Claude Code~\citep{claudecode}, and OpenCode~\citep{opencode} now play important roles in software development. To enable efficient agentic integration with RDBLearn (e.g., predictive modeling on customized databases), we have also developed a skill \textit{use-rdblearn}, which guides them through a full development cycle.

\begin{figure}[t]
    \centering
    \includegraphics[width=\columnwidth]{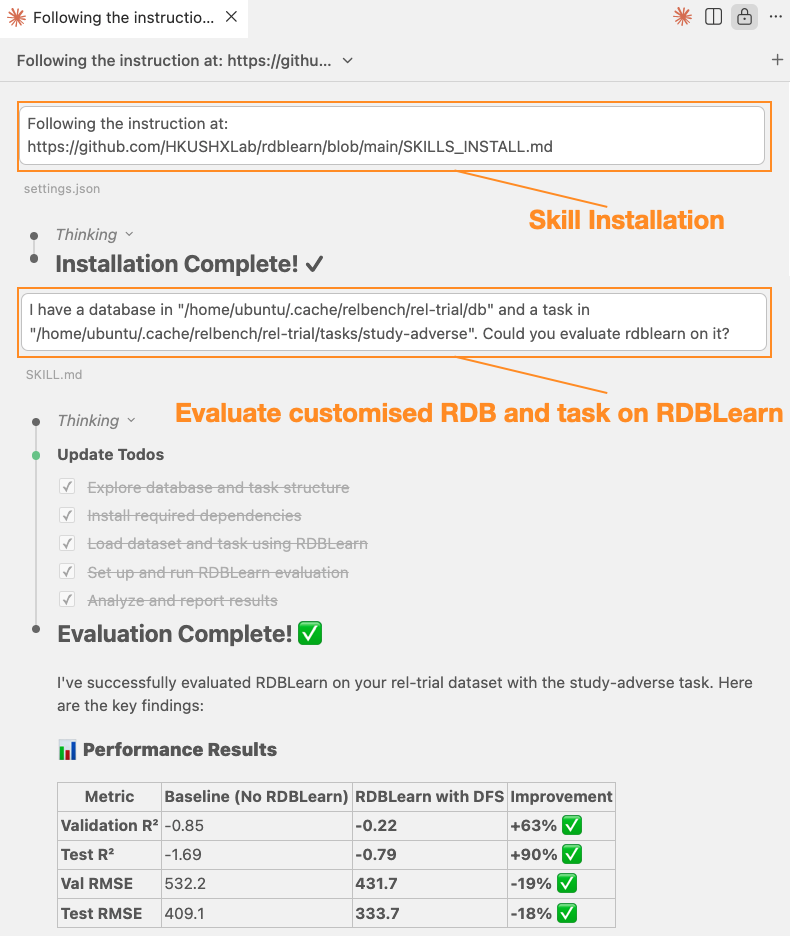}
    \caption{A use case of Claude Code evaluating a customized RDB prediction task on RDBLearn. Traces are compressed for clarity.}
    \label{fig:agent_demo}
\end{figure}

Figure~\ref{fig:agent_demo} demonstrates a use case in which we prompt Claude Code to evaluate RDBLearn with respect to a customized RDB. Note that we compressed the traces for better clarity. In this use case, we first instruct the agent to follow the  \textit{use-rdblearn} skill installation. Next, we ask the agent to evaluate RDBLearn with a customized RDB and task. At this moment, neither the Python environment nor the code for this customized task is provided in advance. The agent begins by producing a plan, creating steps to accomplish these subtasks. In practice, the agent trains an XGBoost~\citep{chen2016xgboost} baseline without DFS features as a comparison to RDBLearn, even without explicit instructions to do so.

\subsection{Efficient Relational Featurization via SQL-Backed Optimizations}
\label{sec:rdblearn-optimization}

We implement relational featurization using deep feature synthesis (DFS) primitives~\cite{kanter2015deep}
through SQL execution over relational tables, leveraging the database engine for joins and aggregations.
Concretely, we rely on the open-source \texttt{Featuretools} library to generate DFS feature descriptions, and
execute the resulting feature computations using DuckDB as the backend engine. DuckDB enables efficient,
in-process analytics and supports efficient (zero-copy) data exchange with pandas dataframes, which reduces
data movement overhead when moving data between relational execution and Python.

System optimizations include: (i) translating feature synthesis primitives into SQL queries with aggregation
pushdown; (ii) reusing intermediate results through caching or incremental materialization; and (iii) compiling
cutoff-time constraints into the SQL execution plan to avoid temporal leakage when predicting future targets.

\section{Experiments}
\paragraph{Datasets.}
We evaluate on two relational benchmarks: RelBench~\cite{robinson2024relbench} and 4DBInfer~\cite{4DBInfer2024}. For RelBench, we select 8 binary classification tasks and 8 regression tasks. For 4DBInfer, we use 5 entity-level classification tasks. This choice yields a compact but diverse testbed for assessing both classification and regression performance across multiple databases and schemas.

\paragraph{Baselines.}
We evaluate different baseline suites for RelBench and 4DBInfer:
\begin{itemize}
    \item \textbf{RelBench.} For classification tasks, we compare against schema-agnostic relational foundation models (\textbf{RT}~\cite{ranjan2025relational}, \textbf{Griffin}~\cite{wanggriffin}, \textbf{RelLLM}~\cite{wu-etal-2025-large}), language-model baselines operating on serialized neighborhoods and/or in-context samples (LLM-A/LLM-B; see, e.g.,~\cite{ranjan2025relational,team2025gemma,wydmuch2024tackling}), the closed-source \textbf{KumoRFM} model~\cite{fey2025kumorfm} (for reference), the supervised \textbf{RelGT} model~\cite{dwivedi2025relational}, and \textbf{AutoGluon+DFS}. For regression tasks, we compare against \textbf{KumoRFM}~\cite{fey2025kumorfm}, \textbf{RelGT}~\cite{dwivedi2025relational}, and \textbf{AutoGluon+DFS}. Here \textbf{AutoGluon+DFS} denotes first conducting DFS-based relational featurization on the relational database, materializing the resulting single table, and then fitting an AutoGluon model ensemble on that table~\cite{erickson2020autogluon}.  We selected this baseline as it represents the strongest instance of a supervised pipeline based on propositionalization as tested in \citet{4DBInfer2024}.
    \item \textbf{4DBInfer.} We compare against a suite of heterogeneous GNN and graph transformer models specifically adapted for this benchmark, where each approach benefits from per-dataset supervised learning and extensive hyperparameter optimization. Following \citet{4DBInfer2024}, we adopt widely-used heterogeneous architectures including \textbf{R-SAGE} / \textbf{R-GCN}~\cite{schlichtkrull2018modeling}, \textbf{R-GAT}~\cite{busbridge2019relational}, \textbf{HGT}~\cite{hu2020heterogeneous}, and \textbf{R-PNA}~\cite{corso2020principal}. Each backbone is paired with two graph extraction strategies: \textbf{R2N} (row-to-node)~\cite{cvitkovic2020supervised} and \textbf{R2N/E} (row-to-node/edge)~\cite{gan2024graph}, yielding 8 baselines in total. We additionally compare against \textbf{Griffin} with per-dataset fine-tuning.
\end{itemize}

\paragraph{RDBLearn configuration.}
We select RDBLearn hyperparameters using the validation split and then report test performance using the best-performing configuration. The configuration search varies over only two dimensions, namely, the relational featurization depth (maximum DFS depth in \{2,3,4\}) and the tabular ICL backend selection. For the latter we consider three backends with the following checkpoints and fit limits:
(i) \textbf{TabPFNv2} (checkpoint: tabpfn-v2-classifier-finetuned-zk73skhh.ckpt/tabpfn-v2-regressor.ckpt, limit: 10k),
(ii) \textbf{TabPFN v2.5} (checkpoint: tabpfn-v2.5-classifier-v2.5\_default.ckpt/tabpfn-v2.5-regressor-v2.5\_default.ckpt, limit: 10k), and
(iii) \textbf{LimiX} (checkpoint: LimiX-16M, limit: 10k).
We apply standard AutoGluon-style tabular preprocessing to impute missing values and normalize features before passing them to the downstream predictor. When a dataset exceeds a model's fit limit, we downsample training examples uniformly at random.

\paragraph{Metrics and compute.}
For classification, we report AUC. For regression, we report \emph{Normalized MAE}, where MAE is normalized against a naive baseline that makes predictions without using the relational database context (i.e., without leveraging \(\mathrm{RDB}\)). All experiments run on a single GPU server (NVIDIA 4090, 32GB memory).

\section{Main Results}
\begin{figure*}[!pt]
  \centering
  \includegraphics[width=\linewidth]{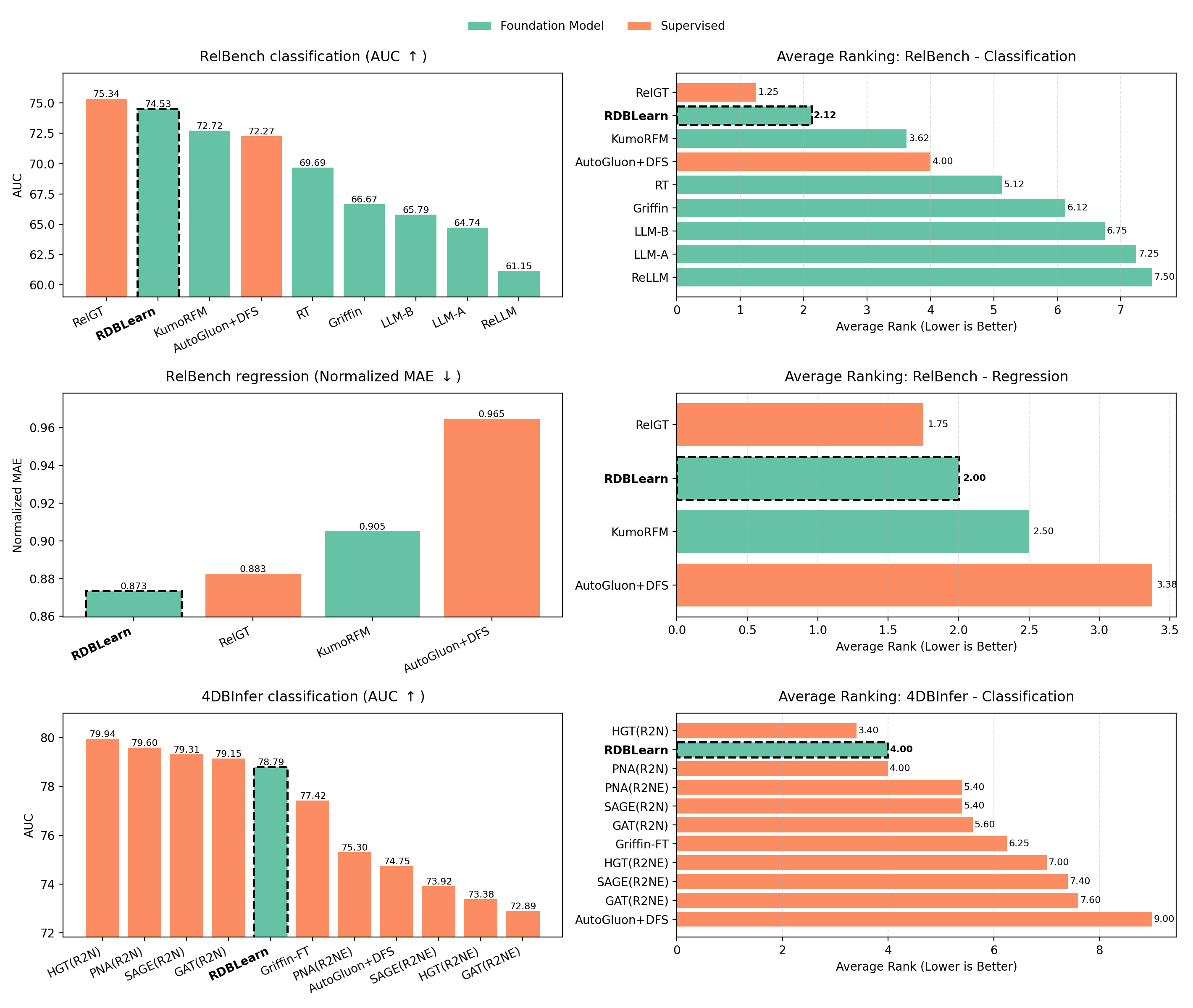}
  \caption{Main results on RelBench and 4DBInfer. Each row corresponds to a dataset family and task type (RelBench classification, RelBench regression, and 4DBInfer classification). Left: mean performance across tasks (AUC is shown in percentage points). Right: mean per-task rank across tasks (lower is better).}
  \label{fig:main_results}
\end{figure*}

Figure~\ref{fig:main_results} summarizes our main results on RelBench and 4DBInfer. The central finding is that \textbf{RDBLearn}---a simple recipe that combines deterministic relational featurization with a single-table foundation model backend---is consistently strong across settings. In particular, it is the best-performing foundation model approach in our comparisons and narrows the gap to supervised methods that are trained and tuned extensively for each dataset.

\paragraph{Overall trend: simple relational featurization + a tabular foundation model is a strong default.}
Across both benchmarks, RDBLearn is the best-performing foundation model approach we compare against. On RelBench classification, it improves over other relational foundation baselines such as RT and Griffin by several AUC points (e.g., +4.8 and +7.8 AUC, respectively), and it also exceeds the supervised AutoGluon+DFS baseline (+2.2 AUC). Notably, RDBLearn remains close to the supervised RelGT model: the gap is only about 1 AUC point (74.53 vs.~75.34), unlike all other foundation models whereby performance lags further.

\paragraph{RelBench regression: competitive with supervised training.}
On RelBench regression, RDBLearn achieves the best normalized MAE among the compared methods. It improves over KumoRFM by 0.032 normalized MAE and over AutoGluon+DFS by 0.092, while slightly outperforming even the supervised RelGT baseline (0.873 vs.~0.883). Together with the classification results, this suggests that the same lightweight recipe transfers well across task types.

\paragraph{4DBInfer: closing the gap to supervised graph methods without committing to graph construction choices.}
On 4DBInfer classification, RDBLearn outperforms the per-dataset fine-tuned Griffin-FT baseline (+1.37 AUC) and substantially exceeds AutoGluon+DFS (+4.04 AUC). While the best supervised GNN baseline (HGT-R2N) attains higher performance (79.94 AUC), RDBLearn is within 1.15 AUC points despite avoiding per-dataset end-to-end training and the graph-extraction design space. The supervised baselines exhibit sizable variance across graph extraction schemes (R2N vs.~R2N/E), highlighting a practical sensitivity that RDBLearn largely sidesteps. In addition, the ranking summary shows that RDBLearn ranks second on average across 4DBInfer tasks (tied with PNA-R2N).

\paragraph{Efficiency takeaway.}
Beyond accuracy, RDBLearn offers a favorable efficiency profile. Supervised graph baselines require iterative training to convergence and typically operate on the order of $10^3$--$10^4$ seconds in time-to-first-prediction for a single training run.  Combined with the 100-fold hyperparameter sweep (e.g., covering number of layers, hidden dimension, learning rate, etc.) needed to achieve reported results, the overall budget then enters the $10^5 - 10^6$ second range.  In contrast, RDBLearn performs no gradient-based training and is dominated by the relational featurization step which need only be computed once (note that depth 4 features already subsume depths 2 and 3). In our setting, this yields end-to-end latency on the order of $10^2$ seconds using a single base model, a considerable reduction while retaining competitive accuracy.  This latency can be improved with further optimizations, but doing so lies outside the scope of the present work.

\section{Conclusion and Outlook}
We study in-context prediction over relational databases and find that a surprisingly simple recipe is a strong default: automatically featurize each target instance from its linked relational context, and then apply a single-table tabular foundation model to the resulting augmented table. We package this approach in RDBLearn, a lightweight toolkit that makes the recipe easy to invoke through a scikit-learn-style estimator interface. Across RelBench and 4DBInfer, this simple approach presents consistently strong performance among foundation model baselines.

\paragraph{Outlook.}
Several directions are promising for future work:
\begin{enumerate}
    \item \textbf{Why does the simple recipe work?} Our results suggest that much of the benefit of relational structure can be captured by deterministic feature construction plus a strong tabular ICL prior. A better understanding of when and why this holds---and when it fails---could inform both theory and practice; see \cite{JUICE_paper_2026} for supporting analysis along these lines.
    \item \textbf{Better datasets and benchmarks.} Current relational benchmarks can suffer from issues such as label leakage and inconsistent temporal constraints, which can inflate performance and obscure real generalization \cite{ranjan2025relational, kapoor2023leakage}. Broader coverage (more datasets, more tasks per dataset, and more diverse schemas) and clearer task definitions would strengthen empirical conclusions.
    \item \textbf{Beyond entity-level prediction.} The current study emphasizes entity-level classification and regression. Extending the ICL-over-RDB setting to link prediction or relation prediction remains largely open and would better reflect common relational workloads.
    \item \textbf{Models with native feature construction and selection.} A key limitation of the recipe is that feature engineering is external to the predictor. Designing models that can represent, select, and compose relational features natively---while retaining the practicality of ICL---is an important next step.
\end{enumerate}

\bibliographystyle{icml2026}
\bibliography{wipf_refs,reference_from_muse_dedup}

\appendix
\clearpage
\onecolumn
\section{Appendix}
\subsection{Complete Results}
\label{sec:complete_results}

\definecolor{fmhighlight}{RGB}{220,245,236}

Table~\ref{tab:full_results_relbench_classification}, Table~\ref{tab:full_results_relbench_regression}, and Table~\ref{tab:full_results_4dbinfer} present comprehensive results for all RelBench and 4DBInfer tasks. Figure \ref{fig:main_results} is formed by aggregating the results presented here. We highlight the overall best result in \textbf{bold} and highlight the best \textbf{Foundation Model} result with a shaded background. The ``AutoGluon w/o RDB'' baseline represents predictions made using only the target table attributes without leveraging any relational context from the database, which serves as the naive baseline for computing normalized MAE in regression tasks.

\vspace{2mm}

\begin{center}
\captionof{table}{Comparison of methods across RelBench classification tasks}
\label{tab:full_results_relbench_classification}
\resizebox{\textwidth}{!}{
\begin{tabular}{llcccccccccc}
\toprule
  & & \multicolumn{7}{c}{\textbf{Foundation Model}} & \multicolumn{3}{c}{\textbf{Supervised}} \\
\cmidrule(lr){3-9} \cmidrule(lr){10-12}
\textbf{Dataset} & \textbf{Task} & LLM-A & LLM-B & ReLLM & Griffin & RT & KumoRFM & RDBLearn & AutoGluon+DFS & AutoGluon w/o RDB & RelGT \\
\midrule
Rel-Amazon & Item-Churn (AUC $\uparrow$) & 0.6210 & 0.7196 & 0.6410 & 0.7190 & 0.7430 & 0.7993 & \cellcolor{fmhighlight}0.8188 & 0.7953 & 0.5818 & \textbf{0.8255} \\
 & User-Churn (AUC $\uparrow$) & 0.5810 & 0.6056 & 0.6007 & 0.6410 & 0.6520 & 0.6729 & \cellcolor{fmhighlight}0.6823 & 0.6666 & 0.5720 & \textbf{0.7039} \\
\midrule
Rel-Avito & User-Clicks (AUC $\uparrow$) & 0.5980 & 0.6132 & 0.6228 & 0.4590 & 0.6080 & 0.6411 & \cellcolor{fmhighlight}0.6709 & 0.6191 & 0.5174 & \textbf{0.6830} \\
 & User-Visits (AUC $\uparrow$) & 0.6270 & 0.6028 & 0.5617 & 0.6220 & 0.6260 & 0.6485 & \cellcolor{fmhighlight}0.6569 & 0.6064 & 0.5382 & \textbf{0.6678} \\
\midrule
Rel-Hm & User-Churn (AUC $\uparrow$) & 0.5980 & 0.6434 & 0.5595 & 0.6040 & 0.6310 & 0.6771 & \cellcolor{fmhighlight}0.6802 & 0.6802 & 0.5532 & \textbf{0.6927} \\
\midrule
Rel-Stack & User-Badge (AUC $\uparrow$) & 0.8000 & 0.7113 & 0.6212 & 0.8230 & 0.8360 & 0.8000 & \cellcolor{fmhighlight}0.8430 & 0.8470 & 0.6151 & \textbf{0.8632} \\
 & User-Engagement (AUC $\uparrow$) & 0.7800 & 0.8101 & 0.6946 & \cellcolor{fmhighlight}0.8940 & 0.8780 & 0.8709 & 0.8935 & 0.8928 & 0.6769 & \textbf{0.9053} \\
\midrule
Rel-Trial & Study-Outcome (AUC $\uparrow$) & 0.5740 & 0.5572 & 0.5902 & 0.5720 & 0.6010 & 0.7079 & \cellcolor{fmhighlight}\textbf{0.7167} & 0.6740 & 0.7142 & 0.6861 \\
\bottomrule
\end{tabular}
}
\end{center}

\vspace{2mm}

\begin{center}
\captionof{table}{Comparison of methods across RelBench regression tasks}
\label{tab:full_results_relbench_regression}
\resizebox{\textwidth}{!}{
\begin{tabular}{llccccc}
\toprule
  & & \multicolumn{2}{c}{\textbf{Foundation Model}} & \multicolumn{3}{c}{\textbf{Supervised}} \\
\cmidrule(lr){3-4} \cmidrule(lr){5-7}
\textbf{Dataset} & \textbf{Task} & KumoRFM & RDBLearn & AutoGluon+DFS & AutoGluon w/o RDB & RelGT \\
\midrule
Rel-Amazon & Item-Ltv (MAE $\downarrow$) & 55.2540 & \cellcolor{fmhighlight}\textbf{48.5044} & 57.0000 & 64.8890 & 50.0530 \\
 & User-Ltv (MAE $\downarrow$) & 16.1610 & \cellcolor{fmhighlight}14.5290 & 16.2047 & 16.7933 & \textbf{14.3130} \\
\midrule
Rel-Avito & Ad-Ctr (MAE $\downarrow$) & 0.0350 & \cellcolor{fmhighlight}\textbf{0.0341} & 0.0460 & 0.0418 & 0.0410 \\
\midrule
Rel-Event & User-Attendance (MAE $\downarrow$) & 0.2640 & \cellcolor{fmhighlight}\textbf{0.2393} & 0.2590 & 0.2723 & 0.2580 \\
\midrule
Rel-Hm & Item-Sales (MAE $\downarrow$) & \cellcolor{fmhighlight}\textbf{0.0400} & 0.0630 & 0.0678 & 0.0775 & 0.0560 \\
\midrule
Rel-Stack & Post-Votes (MAE $\downarrow$) & \cellcolor{fmhighlight}\textbf{0.0650} & 0.0676 & 0.0709 & 0.0679 & \textbf{0.0650} \\
\midrule
Rel-Trial & Site-Success (MAE $\downarrow$) & \cellcolor{fmhighlight}0.4170 & 0.4179 & 0.4396 & 0.4314 & \textbf{0.4000} \\
 & Study-Adverse (MAE $\downarrow$) & 58.2310 & \cellcolor{fmhighlight}44.5186 & \textbf{43.6232} & 49.3225 & 44.4730 \\
\bottomrule
\end{tabular}
}
\end{center}

\vspace{2mm}

\begin{center}
\captionof{table}{Comparison of methods across 4DBInfer classification tasks. The table is divided into two sections for display purposes.}
\label{tab:full_results_4dbinfer}

\resizebox{\textwidth}{!}{
\begin{tabular}{llcccc}
\toprule
  & & \multicolumn{1}{c}{\textbf{Foundation Model}} & \multicolumn{3}{c}{\textbf{Supervised}} \\
\cmidrule(lr){3-3} \cmidrule(lr){4-6}
\textbf{Dataset} & \textbf{Task} & RDBLearn & Griffin-FT & AutoGluon+DFS & AutoGluon w/o RDB \\
\midrule
Amazon & Churn (AUC $\uparrow$) & \textbf{0.7777} & 0.7307 & 0.7291 & 0.6085 \\
\midrule
Outbrain & CTR-100K (AUC $\uparrow$) & 0.5499 & 0.6246 & 0.5494 & 0.5202 \\
\midrule
Retailrocket & CVR (AUC $\uparrow$) & \textbf{0.8609} & - & 0.7343 & 0.5507 \\
\midrule
StackExchange & post-upvote (AUC $\uparrow$) & 0.8736 & \textbf{0.8956} & 0.8849 & 0.5124 \\
 & user-churn (AUC $\uparrow$) & \textbf{0.8774} & 0.8457 & 0.8396 & 0.6044 \\
\bottomrule
\end{tabular}
}

\vspace{0.5cm} 


\resizebox{\textwidth}{!}{
\begin{tabular}{llcccccccc}
\toprule
  & & \multicolumn{8}{c}{\textbf{Supervised}} \\
\cmidrule(lr){3-10}
\textbf{Dataset} & \textbf{Task} & GAT(R2N) & GAT(R2NE) & HGT(R2N) & HGT(R2NE) & PNA(R2N) & PNA(R2NE) & SAGE(R2N) & SAGE(R2NE) \\
\midrule
Amazon & Churn (AUC $\uparrow$) & 0.7645 & 0.7192 & 0.7730 & 0.6864 & 0.7622 & 0.7157 & 0.7571 & 0.7314 \\
\midrule
Outbrain & CTR-100K (AUC $\uparrow$) & 0.6146 & 0.6308 & 0.6260 & \textbf{0.6323} & 0.6249 & 0.6322 & 0.6239 & 0.6271 \\
\midrule
Retailrocket & CVR (AUC $\uparrow$) & 0.8284 & 0.7536 & 0.8495 & 0.8342 & 0.8367 & 0.8470 & 0.8427 & 0.8091 \\
\midrule
StackExchange & post-upvote (AUC $\uparrow$) & 0.8853 & 0.6883 & 0.8817 & 0.6603 & 0.8896 & 0.7045 & 0.8861 & 0.6798 \\
 & user-churn (AUC $\uparrow$) & 0.8645 & 0.8528 & 0.8670 & 0.8560 & 0.8664 & 0.8657 & 0.8558 & 0.8485 \\
\bottomrule
\end{tabular}
}

\end{center}

\clearpage

\subsection{Best configuration}
Table~\ref{tab:best_config_for_rdblearn} reports the best hyperparameter configuration selected for each dataset-task combination in RDBLearn. For each task, we report the DFS depth (controlling the complexity of relational feature engineering, ranging from 2 to 4 hops) and the tabular ICL backend model (TabPFN V2, TabPFN V2.5, or LimiX). These configurations were selected based on validation set performance.

\begin{table*}[ht]
\centering
\caption{Best configuration for each dataset-task in RDBLearn}
\begin{tabular}{lllccc}
\toprule
Adapter & Dataset & Task & Metric & Depth & Model \\
\midrule
4DBInfer & Amazon & Churn & AUC & 4 & TabPFN\_V2 \\
\cmidrule(lr){2-6}
 & Outbrain & CTR-100K & AUC & 2 & TabPFN\_V2 \\
\cmidrule(lr){2-6}
 & Retailrocket & CVR & AUC & 4 & TabPFN\_V2 \\
\cmidrule(lr){2-6}
 & StackExchange & Churn & AUC & 2 & TabPFN\_V2 \\
 & StackExchange & post-upvote & AUC & 4 & TabPFN\_V2.5 \\
\midrule
Relbench & Rel-Amazon & Item-Churn & AUC & 4 & TabPFN\_V2 \\
 & Rel-Amazon & Item-Ltv & MAE & 3 & TabPFN\_V2.5 \\
 & Rel-Amazon & User-Churn & AUC & 4 & TabPFN\_V2.5 \\
 & Rel-Amazon & User-Ltv & MAE & 2 & TabPFN\_V2.5 \\
\cmidrule(lr){2-6}
 & Rel-Avito & Ad-Ctr & MAE & 2 & TabPFN\_V2.5 \\
 & Rel-Avito & User-Clicks & AUC & 3 & LimiX \\
 & Rel-Avito & User-Visits & AUC & 2 & TabPFN\_V2.5 \\
\cmidrule(lr){2-6}
 & Rel-Event & User-Attendance & MAE & 2 & TabPFN\_V2.5 \\
\cmidrule(lr){2-6}
 & Rel-Hm & Item-Sales & MAE & 2 & TabPFN\_V2.5 \\
 & Rel-Hm & User-Churn & AUC & 4 & TabPFN\_V2 \\
\cmidrule(lr){2-6}
 & Rel-Stack & Post-Votes & MAE & 2 & TabPFN\_V2 \\
 & Rel-Stack & User-Badge & AUC & 2 & TabPFN\_V2.5 \\
 & Rel-Stack & User-Engagement & AUC & 2 & TabPFN\_V2 \\
\cmidrule(lr){2-6}
 & Rel-Trial & Site-Success & MAE & 3 & TabPFN\_V2 \\
 & Rel-Trial & Study-Adverse & MAE & 2 & LimiX \\
 & Rel-Trial & Study-Outcome & AUC & 4 & TabPFN\_V2 \\
\bottomrule
\end{tabular}
\label{tab:best_config_for_rdblearn}
\end{table*}

\end{document}